\documentclass[aps,twocolumn,prl,superscriptaddress,showpacs,floatfix]{revtex4}
\usepackage{epsfig, amssymb}
\usepackage[dvips]{pstcol}

\begin{document}
\advance\hoffset by  -4mm

\def\cal{\mathcal}
\def\piz{\pi ^0 }
\def\pip{\pi ^+ }
\def\pim{\pi ^- }
\def\BBbar{B\overline{B}}
\def\epem{e^+ e^-}
\def\Elsum{E_{\ell ^+}+E_{\ell ^-}}
\def\Emiss{E_{\rm miss}}
\def\pmiss{\vec{p}_{\rm miss}}
\def\Mmiss{M_{\rm miss}}
\def\GeV{{\rm GeV}}
\def\GeVc{{{\rm GeV}/c}}
\def\GeVcc{{{\rm GeV}/c^2}}
\def\MeV{{\rm MeV}}
\def\MeVc{{{\rm MeV}/c}}
\def\MeVcc{{\rm MeV/}c^2}
\def\to{\rightarrow}
\def\BR{{\cal B}}

\def\gevc{GeV/$c$}
\def\mevc{MeV/$c$}
\def\gevcc{GeV/$c^2$}
\def\mevcc{MeV/$c^2$}
\def\ra{\rightarrow}
\def\to{\rightarrow}
\newcommand{\rt}{\rightarrow}
\newcommand{\cont}{e^+e^- \rt q\overline{q}}
\newcommand{\mupmum}{\mu^+ \mu^-}
\newcommand{\pippim}{\pi^+ \pi^-}
\newcommand{\mmppx}{$(\mu^+ \mu^- \pi^+ \pi^- X)$}
\newcommand{\mmumu}{$M_{\mu\mu}$}
\newcommand{\dmtot}{$M_{\mu\mu\pi\pi}-M_{\mu\mu}$}
\newcommand{\ups}{$\Upsilon$}
\newcommand{\uIs}{$\Upsilon(1S)$}
\newcommand{\uIIs}{$\Upsilon(2S)$}
\newcommand{\uIIIs}{$\Upsilon(3S)$}
\newcommand{\utts}{$\Upsilon(2S,3S)$}
\newcommand{\uIVs}{$\Upsilon(4S)$}
\newcommand{\uIsd}{$\Upsilon(1S) \rightarrow \mu^{+} \mu^{-} $}
\newcommand{\uttsd}{$\Upsilon(2S,3S)\rightarrow\Upsilon(1S) \pi^{+} \pi^{-} $}
\newcommand{\uIVsd}{$\Upsilon(4S) \rightarrow \Upsilon(1S) \pi^{+} \pi^{-} $}
\newcommand{\uIVsdt}{$\Upsilon(4S)\to \Upsilon(1S)\pi^+\pi^- \to \mu^+ \mu^-\pi^+\pi^-$}
\newcommand{\brtot}{$\mathcal{B}(\Upsilon(4S)\to \Upsilon(1S)\pi^+\pi^-)\, = \,
(0.81 \,\pm\, 0.12(\mathrm{stat.})\, \pm \,0.08(\mathrm{sys.}))\times 10^{-4}$}
\title{\Large \rm  Measurement of the branching fraction  for the decay $\Upsilon(4S) \ra \Upsilon(1S) \pi^{+} \pi^{-} $}
\date{\today}
\affiliation{Budker Institute of Nuclear Physics, Novosibirsk}
\affiliation{Chiba University, Chiba}
\affiliation{University of Cincinnati, Cincinnati, Ohio 45221}
\affiliation{T. Ko\'{s}ciuszko Cracow University of Technology, Krakow}
\affiliation{Department of Physics, Fu Jen Catholic University, Taipei}
\affiliation{Justus-Liebig-Universit\"at Gie\ss{}en, Gie\ss{}en}
\affiliation{The Graduate University for Advanced Studies, Hayama}
\affiliation{Hanyang University, Seoul}
\affiliation{University of Hawaii, Honolulu, Hawaii 96822}
\affiliation{High Energy Accelerator Research Organization (KEK), Tsukuba}
\affiliation{Hiroshima Institute of Technology, Hiroshima}
\affiliation{Institute of High Energy Physics, Chinese Academy of Sciences, Beijing}
\affiliation{Institute of High Energy Physics, Vienna}
\affiliation{Institute of High Energy Physics, Protvino}
\affiliation{Institute for Theoretical and Experimental Physics, Moscow}
\affiliation{J. Stefan Institute, Ljubljana}
\affiliation{Kanagawa University, Yokohama}
\affiliation{Korea University, Seoul}
\affiliation{Kyungpook National University, Taegu}
\affiliation{\'Ecole Polytechnique F\'ed\'erale de Lausanne (EPFL), Lausanne}
\affiliation{University of Maribor, Maribor}
\affiliation{University of Melbourne, School of Physics, Victoria 3010}
\affiliation{Nagoya University, Nagoya}
\affiliation{Nara Women's University, Nara}
\affiliation{National Central University, Chung-li}
\affiliation{National United University, Miao Li}
\affiliation{Department of Physics, National Taiwan University, Taipei}
\affiliation{H. Niewodniczanski Institute of Nuclear Physics, Krakow}
\affiliation{Nippon Dental University, Niigata}
\affiliation{Niigata University, Niigata}
\affiliation{University of Nova Gorica, Nova Gorica}
\affiliation{Novosibirsk State University, Novosibirsk}
\affiliation{Osaka City University, Osaka}
\affiliation{Panjab University, Chandigarh}
\affiliation{Saga University, Saga}
\affiliation{University of Science and Technology of China, Hefei}
\affiliation{Seoul National University, Seoul}
\affiliation{Sungkyunkwan University, Suwon}
\affiliation{University of Sydney, Sydney, New South Wales}
\affiliation{Tata Institute of Fundamental Research, Mumbai}
\affiliation{Toho University, Funabashi}
\affiliation{Tohoku Gakuin University, Tagajo}
\affiliation{Tohoku University, Sendai}
\affiliation{Department of Physics, University of Tokyo, Tokyo}
\affiliation{Tokyo Institute of Technology, Tokyo}
\affiliation{Tokyo Metropolitan University, Tokyo}
\affiliation{Tokyo University of Agriculture and Technology, Tokyo}
\affiliation{IPNAS, Virginia Polytechnic Institute and State University, Blacksburg, Virginia 24061}
\affiliation{Yonsei University, Seoul}
  \author{A.~Sokolov}\affiliation{Institute of High Energy Physics, Protvino} 
  \author{M.~Shapkin}\affiliation{Institute of High Energy Physics, Protvino} 
  \author{H.~Aihara}\affiliation{Department of Physics, University of Tokyo, Tokyo} 
  \author{K.~Arinstein}\affiliation{Budker Institute of Nuclear Physics, Novosibirsk}\affiliation{Novosibirsk State University, Novosibirsk} 
  \author{T.~Aushev}\affiliation{\'Ecole Polytechnique F\'ed\'erale de Lausanne (EPFL), Lausanne}\affiliation{Institute for Theoretical and Experimental Physics, Moscow} 
  \author{A.~M.~Bakich}\affiliation{University of Sydney, Sydney, New South Wales} 
  \author{E.~Barberio}\affiliation{University of Melbourne, School of Physics, Victoria 3010} 
  \author{A.~Bay}\affiliation{\'Ecole Polytechnique F\'ed\'erale de Lausanne (EPFL), Lausanne} 
  \author{K.~Belous}\affiliation{Institute of High Energy Physics, Protvino} 
  \author{V.~Bhardwaj}\affiliation{Panjab University, Chandigarh} 
% ...
\collaboration{The Belle Collaboration}

\begin{abstract}
We study transitions between {\ups} states with the emission of
charged pions using 604.6~fb$^{-1}$ of data collected
with the Belle detector at the KEKB asymmetric-energy $e^+ e^-$ collider.
The measured product branching fraction is
$\mathcal{B}(\Upsilon(4S)\to \Upsilon(1S)\pi^+\pi^-)\times
\mathcal{B}(\Upsilon(1S)\to \mu^+\mu^-)=
{\color{black}{(2.11 \pm 0.30(\mathrm{stat.})\pm 0.14(\mathrm{sys.}))}}
\times 10^{-6}$
and the partial decay width is 
$\Gamma(\Upsilon(4S) \rightarrow \Upsilon(1S) \pi^{+} \pi^{-})
\,=\,{\color{black}{(1.75\, \pm \,0.25(\mathrm{stat.})\, 
\pm \,0.24(\mathrm{sys.}))}}$~keV.
\end{abstract}
\maketitle

%
The bottomonium state $\Upsilon(4S)$ has a mass above the
threshold for  $B \overline{B}$ pair
production and decays mainly into $B$-meson pairs
($\mathcal{B}(\Upsilon(4S) \to B\overline{B})\,>\,96\%$~\cite{PDG}). 
Recently, the decay modes
$\Upsilon(4S) \rightarrow \Upsilon(mS) \pi \pi$ with $m\,=\, 1, 2$
{{\color{black} {as well as $\Upsilon(4S) \rightarrow \eta 
\Upsilon(1S)$~\cite{etab}}}
have been also observed.
{{\color{black} {These decays as well as the anomalously large width of the
$\Upsilon(5S) \rightarrow \Upsilon(mS) \pi \pi$ ($m\,=\, 1, 2, 3)$
 transitions discovered by Belle~\cite{u5s}}} give additional information 
about various  QCD models 
that are used to describe hadronic transitions 
of heavy quarkonia~\cite{theory}.
Preliminary evidence for the decay
$\Upsilon(4S) \rightarrow \Upsilon(1S) \pi^+ \pi^-$ was presented  
by the Belle Collaboration in Ref.~\cite{prel}.
The {\it BABAR} Collaboration reported measurements of the 
$\Upsilon(4S)$ transition to the $\Upsilon(1S)$ or $\Upsilon(2S)$
with the emission of a $\pi^+ \pi^-$ pair~\cite{BaBar,etab}.   
The first Belle measurements of
$\Upsilon(4S) \rightarrow \Upsilon(1S) \pi^+ \pi^-$ were published in
Ref.~\cite{4S}. 
The product branching fraction 
$\mathcal{B}(\Upsilon(4S)\to \Upsilon(1S)\pi^+\pi^-)\times
\mathcal{B}(\Upsilon(1S)\to \mu^+\mu^-)$ from Belle differs by 2.4 standard 
deviations from
the correponding value from {\it BABAR}~\cite{etab}.
In this paper we present 
a new study of the decay mode
 $\Upsilon(4S) \rightarrow \Upsilon(1S) \pi^+ \pi^-$
from the Belle experiment using a larger data sample
and relaxed signal selection criteria.

We use 604.6~fb$^{-1}$ of data collected on the $\Upsilon(4S)$ resonance
with the Belle detector~\cite{Belle} at the KEKB asymmetric-energy 
$e^+ e^-$ collider~\cite{KEKB}.
We study
$\Upsilon(4S) \rightarrow \Upsilon(1S) \pi^+ \pi^-$ decays with
a subsequent {\uIsd} transition.
Charged particles are reconstructed and identified
in the Belle detector, which
consists of a silicon vertex detector (SVD),
central drift chamber (CDC),
aerogel threshold Cherenkov counters (ACC),
time-of-flight (TOF) scintillation counters,
an electromagnetic calorimeter (ECL),
and a $K_L$-muon detector (KLM).

Charged tracks must originate from within a region of radius $1\,{\rm cm}$ 
and axial length $\pm 5\,{\rm cm}$ centered in the $e^+ e^-$ 
interaction point and not be associated 
with a well-reconstructed $K_S^0$ meson, $\Lambda$ baryon, or converted photon;
each charged track should have a momentum
transverse to the beam axis ($p_T$)
of greater than 0.1~GeV/$c$.
Charged particles are assigned
a likelihood  $\mathcal{L}_i$~\cite{MUID} ($i$ = $\mu$, $\pi$, $K$)
 based on the matching of hits in
 	the KLM to the track extrapolated from the CDC, and identified as
 	muons if the likelihood ratio $P_{\mu}=\mathcal{L}_{\mu}/
(\mathcal{L}_{\mu}+\mathcal{L}_{\pi}+\mathcal{L}_K)$ exceeds  0.8,
corresponding to a muon detection efficiency of approximately 91.5\%
over the polar angle range
$20^{\circ} \, \le  \,\theta \, \le \, 155^{\circ}$
and the momentum range
$0.7\, \mathrm{GeV}/c \,\le \, p  \,\le \,3.0\,\mathrm{GeV}/c$
in the laboratory frame. Electron identification uses a similar likelihood
ratio $P_e$~\cite{EID} based on CDC, ACC, and ECL information.
Charged particles that are not identified as muons
and have a likelihood ratio $P_e\,< \,$0.1 are treated as pions.
Calorimeter clusters not associated with reconstructed charged tracks
and with energies greater than 50~MeV
are classified as photon candidates.

Candidates for $\Upsilon(4S) \rightarrow \Upsilon(1S) \pi^+ \pi^-$
decays  with the subsequent
leptonic decay $\Upsilon(1S)\to \mu^+ \mu^-$
are selected from the standard hadronic-event sample
for the first $492\,\rm fb^{-1}$ data set (sample I), 
while an additional $\tau$-enriched sample is also
used for the remaining  $113\,\rm fb^{-1}$ data set (sample II).
The relevant selection criteria for the standard hadronic-event sample 
are the following: three or more charged tracks; 
a visible energy $E_{\rm vis}$ of at least $0.2\sqrt{s}$, where $\sqrt{s}$
is  the center-of-mass (c.m.) energy;
 a calorimeter energy deposit in the range 
$0.1 \,\le \,E_{\rm sum}/\sqrt{s}\,\le \,0.8$;
 and a maximum of $0.5\sqrt{s}$ for the magnitude of $P_{\rm z}$,
 the sum of the $z$ components of the momenta of each charged track 
and neutral cluster, where the $z$ axis is defined to be the direction 
opposite to the positron beam.
The variables $E_{\rm vis}$,  $E_{\rm sum}$, and $P_{\rm z}$
are evaluated in the c.m.\ system.

Unfortunately, the hadronic event sample  criterion
$E_{\rm sum}/\sqrt{s}\,\ge \,0.1$ rejects a considerable fraction of our 
signal events.
However, some of these lost events are recovered in the 
$\tau$-enriched sample because it 
has  a loose constraint on the $E_{\rm sum}$ variable 
($E_{\rm sum}\,\le \,10\,$GeV).   
Only the newest $\tau$-enriched data are used because the earlier subset 
included a requirement
on the sum of the magnitudes of the charged 
track momenta 
in the c.m. frame (below $10\,{\rm GeV}/c$) that rejected most of our signal. 
Other selection criteria for the $\tau$-enriched sample{{\color{black}{,
while not critical for this study, are enumerated below. 
The number of charged tracks in an event  
should be greater than one and less than nine with zero net charge. 
 The maximum $p_T$ among the tracks is 
required to be greater than 0.5~GeV/$c$.
Beam-related background is rejected by requiring that the position 
of the reconstructed event vertex be less than 1~cm from the
interaction point (IP)
in the transverse direction and less than 3~cm from the IP
along the beam direction.
To suppress background from Bhabha and $\mu^+\mu^-$ events, 
the maximum opening angle between charge tracks is required to be less
than $175^{\circ}$ in the CM frame. 
}}}

To select $\Upsilon(4S) \to \Upsilon(1S) \pi^+ \pi^-$ decays, 
an event is required to contain exactly four charged tracks 
with a $\mu^+ \mu^-$ pair having an invariant
 mass $M_{\mu\mu}$ above $9.0\,{\rm GeV}/c^2$
and a $\pi^+ \pi^-$ pair whose opening angle $\theta_{\pi\pi}$ 
in the laboratory frame satisfies $\cos\theta_{\pi\pi} < 0.95$. 
The latter criterion suppresses the radiative return process~\cite{rad} 
$e^+ e^- \to \Upsilon(1S)\gamma$ 
{{\color{black} {as well as $e^+ e^- \to \mu^+ \mu^- \gamma$}}}
wherein the photon converts to 
an $e^+ e^-$ pair that is misidentified as a pion pair. 
Poorly reconstructed events are discarded by requiring a visible energy 
in the laboratory frame of 
$10.5\,{\rm GeV} < E_{\rm vis}^{\rm lab} < 12.5\,{\rm GeV}$. 

To identify parent resonances that decay into the 
$\Upsilon(1S) \ \pi^+ \pi^-$ final state,
the distribution of $M_{\mu\mu}$ vs. the mass difference
$\Delta M = M_{\mu\mu\pi\pi} - M_{\mu\mu}$
is examined (see Fig.~\ref{fig1}) for the selected data sample.
The cluster of events in the  parallelogram centered 
at ($\Delta M, M_{\mu\mu}$)
= (1.12, 9.46) GeV/$c^2$ is from the transition
$\Upsilon(4S)\to \Upsilon(1S)\pi^+\pi^-$.
 The other clusters are  due to the decays
$\Upsilon(2S) \rightarrow \Upsilon(1S) \pi^+ \pi^-$ and
$\Upsilon(3S) \rightarrow \Upsilon(1S) \pi^+ \pi^-$, where
the $\Upsilon(2S)$ and $\Upsilon(3S)$ are produced  predominantly 
by radiative return i.e.
$e^+e^- \rightarrow \Upsilon(mS) \gamma$.
\begin{figure}
\includegraphics[width=0.4\textwidth] {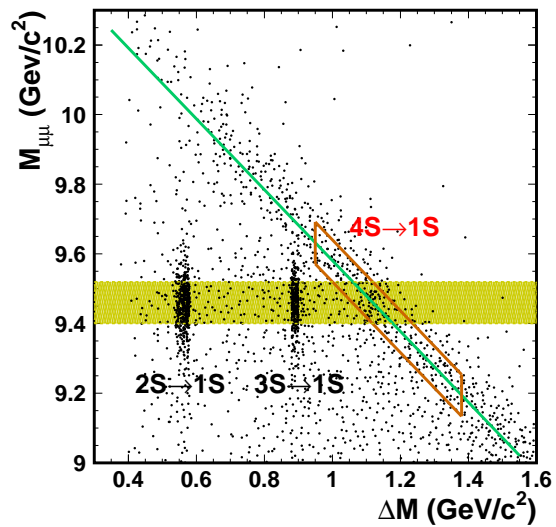}
\caption{The $M_{\mu\mu}$ {\it vs.} $\Delta M$ distribution for the candidate events. 
The $\pm$60-MeV high horizontal shaded band is centered 
on the nominal $\Upsilon(1S)$ mass. 
The clusters on the lower left correspond to  
$\Upsilon(2S) \to \Upsilon(1S) \pi^+ \pi^-$ and 
$\Upsilon(3S) \to \Upsilon(1S) \pi^+\pi^-$ transitions. 
The diagonal line indicates the kinematic boundary 
$M_{\mu\mu\pi\pi} = \sqrt{s}$. 
The parallelogram straddling this line defines the fitting region for 
$\Upsilon(4S) \to \Upsilon(1S) \pi^+ \pi^-$ candidates.
}
\label{fig1}
\end{figure}

\begin{figure}
\includegraphics[width=0.4\textwidth] {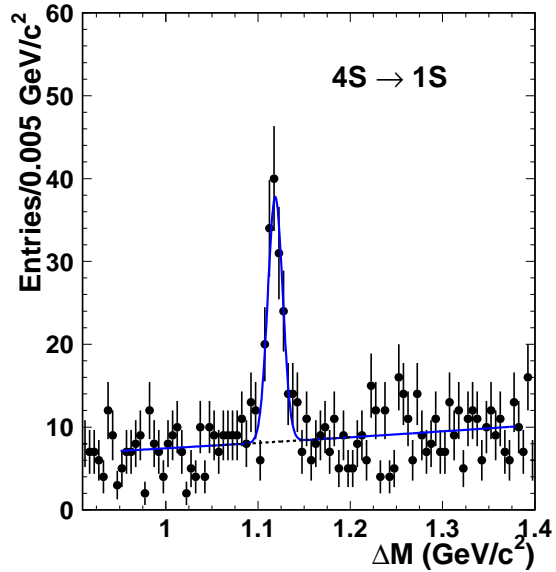}
\caption{The fit to the $\Delta M$ distribution  
for events within the parallelogram of Fig. 1 
using a Gaussian for the signal
and a second-order polynomial for the background (dotted line).
The solid curve shows the sum of the Gaussian and the polynomial function.
}
\label{fig2}
\end{figure}

The rightmost cluster in Fig. 1 contains events from the process 
$\Upsilon(4S) \to \Upsilon(1S) \pi^+ \pi^-$. 
The dominant background processes, 
$e^+ e^-  \rightarrow \mu^+ \mu^- \gamma$ ($\gamma  \rightarrow e^+ e^-$),
$e^+ e^-  \rightarrow \mu^+ \mu^-  \mu^+ \mu^- $, and 
$e^+e^- \rightarrow \mu^+ \mu^- \pi^+\pi^-$,
accumulate at the kinematic boundary indicated by the diagonal line 
in Fig.~1.
To capture the $\Upsilon(4S) \to \Upsilon(1S) \pi^+ \pi^-$ signal 
as well as estimate this background more reliably, we fit the distribution 
of $\Delta M$ for events within the parallelogram of Fig.~1, 
whose boundaries correspond to 
$|M_{\mu\mu\pi\pi} - \sqrt{s}| < 60\,{\rm MeV}/c^2$ and
 $950\,{\rm MeV}/c^2 < \Delta M < 1380\,{\rm MeV}/c^2$. 
The fit, shown in Fig. 2, includes a Gaussian for the signal 
and a quadratic function for the background. The fitted Gaussian is 
centered at $(1118.7 \,\pm \, 1.2)\,\mathrm{MeV}/c^2$,
which is in good agreement with the nominal 
$m_{\Upsilon(4S)}-m_{\Upsilon(1S)}$ mass difference, 
and has a width of $(8.1\,\pm \, 1.0)\,\mathrm{MeV}/c^2$, 
which is consistent with the detector's estimated
$\Delta M$ resolution.
The signal yield in the interval
$1105\,\mathrm{MeV}/c^2\,<\,\Delta M \,< \,1135\,\mathrm{MeV}/c^2$,
 determined as the difference of the number of events and the 
fitted background 
in this interval, is $N_{\rm ev} \,=  \,113.7 \,\pm  \,16.3$,
with a statistical significance of $11.2\,\sigma$,
corresponding to
${-2{\rm ln}(\mathcal{L}_0/\mathcal{L}_{\rm max})}=135.0$
with three fit parameters (mass, width, and yield).
Here, $\mathcal{L}_0$ and
$\mathcal{L}_{\rm max}$ are the likelihood values returned by the fit with
the signal yield fixed at zero and its best fit value,
respectively.

Additional information can be obtained from the study of the
$\pi^+ \pi^-$ system.
{\color{black} {Background-subtracted and efficiency-corrected 
distribution of $\pi^+ \pi^-$ invariant mass ($M_{\pi\pi}$)}} 
is shown in Fig. 3 for events within the signal subregion 
($1105\,\mathrm{MeV}/c^2\,<\,\Delta M \,< \,1135\,\mathrm{MeV}/c^2$)
of the parallelogram in Fig.~1. 
(The background is estimated from the sideband subregion 
$950\,\mathrm{MeV}/c^2\,<\,\Delta M \,< \,1075\,\mathrm{MeV}/c^2$ and
$1175\,\mathrm{MeV}/c^2\,<\,\Delta M \,< \,1350\,\mathrm{MeV}/c^2$).
The EvtGen event generator~\cite{EvtGen} 
with a matrix element that accounts for particle spins~\cite{Brown},
is used  to produce {\uIVsdt} events
that are then  passed through the detector
simulation~\cite{geant3} and reconstruction programs.

\begin{figure}
  \includegraphics[width=0.4\textwidth] {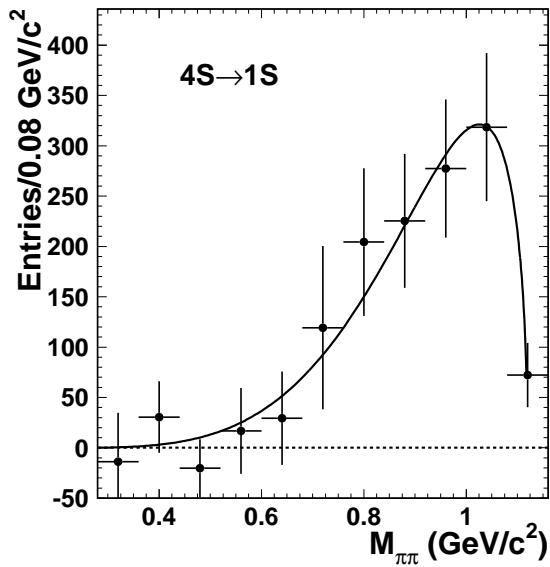}
\caption{{\color{black} {Background-subtracted and efficiency-corrected}} 
distribution of $\pi^+ \pi^-$ invariant mass {\color{black} {($M_{\pi\pi}$)}} 
for events within the signal
subregion of the parallelogram in Fig. 1. The solid curve
shows the $M_{\pi\pi}$ distribution predicted
by the models of Ref.~\cite{Brown}.}
\label{fig3}
\end{figure}

The $M_{\pi\pi}$ distribution in Fig. 3 can be described 
using the shape predicted
by the models of Ref.~\cite{Brown},
in which suppression of small $\pi^+\pi^-$
invariant masses follows from partial conservation of axial current.
The goodness of fit, for 10 degrees of freedom (NDF), is
$\chi^2/{\rm NDF}= 0.35$.

The branching fraction for the $\Upsilon(4S)\to \Upsilon(1S)\pi^+\pi^-$ 
decay is determined from 
$\mathcal{B}(\Upsilon(4S) \to \Upsilon(1S)\pi^+\pi^-)\, =\,
N_{\rm ev}/(N_{\Upsilon(4S)}
\cdot \varepsilon \cdot \mathcal{B}(\Upsilon(1S) \rightarrow \mu^+ \mu^-))$, 
where $N_{\rm ev}$ is the extracted signal yield, 
$N_{\Upsilon(4S)}$ is the estimated number of $\Upsilon(4S)$ events produced, 
$\varepsilon$ is the signal detection efficiency (calculated separately 
for samples I and II), and 
$\mathcal{B}(\Upsilon(1S) \rightarrow \mu^+ \mu^-)\,=\,
(2.48\,\pm\, 0.05)\%$ is the PDG-tabulated 
branching fraction for the daughter decay. The efficiencies are 
calculated from Monte Carlo simulations. 
For the hadronic-event simulation in sample I,
we apply a correction to $E_{\rm sum}/\sqrt{s}$, 
one of the variables used to select hadronic events, so that 
this distribution agrees with that of the data. This correction 
is also applied to  sample II, where it changes 
the efficiency by a few per cent. The results are given in Table I.
\begin{table}
\caption{Total number of  $\Upsilon(4S)$ ($N_{\Upsilon(4S)}$), signal yield ($N_{\rm ev}$),
reconstruction efficiency ($\varepsilon$), and branching fraction (${\cal B}$) for the 
$\Upsilon(4S)\to \Upsilon(1S)\pi^+\pi^-$ decay.}
\medskip
\label{simulfit}
\begin{tabular}{ccccc}\hline\hline
Data  sample & $N_{\Upsilon(4S)}$, $10^{6}$ & $N_{\rm ev}$ & $\varepsilon$, 
\% & ${\cal B}$, $10^{-4}$ \\\hline
 I & 534.6$\pm$7.0 & 52.2$\pm$10.7 & {\color{black}{4.5}} & 
{\color{black}{0.86$\pm$0.18}}\\
II & 122.1$\pm$1.4 & 61.3$\pm$12.1 & {\color{black}{25.1}} & 
{\color{black}{0.84$\pm$0.17}}\\\hline\hline
\end{tabular}
\end{table}

The systematic error in the reconstruction efficiency
due to the last correction is 8\% for sample I
 and essentially zero for  sample II.
The systematic uncertainty in the reconstruction efficiency
due to lack of knowledge  of the {\uIVsdt}
decay matrix element
is estimated by comparing the parameterization of the $M_{\pi\pi}$
distribution in the models of Ref.~\cite{Brown} and in a phase space model.
We estimate this systematic uncertainty   as  half 
of the variation in the efficiency, and it is equal to 2.0\%(3.1\%);
here and below, the first(second) value gives the systematic uncertainty 
for sample I (sample II).
The signal yield is extracted by an unbinned extended maximum 
likelihood fit to the $\Delta M$
distribution for events in the parallelogram (Fig.1) 
using a Gaussian for the signal
and a second-order polynomial for the background.
The signal yield in the signal interval
$1105\,\mathrm{MeV}/c^2\,<\,\Delta M \,< \,1135\,\mathrm{MeV}/c^2$
is determined as the difference of the number of events and the 
fitted background in this interval.
The signal yield from the fitted Gaussian area has a larger statistical error.
The systematic uncertainties from the discrepancies between these 
two evaluations of the signal yield are 2.2\% and 1.2\%
for samples I and II, respectively.
Other systematic uncertainties come from the
choice of the fit range (0.3\%, 2.5\%),
the choice of the signal range (2\%, 0.6\%),
the choice of the signal box width  (2.4\%, 1.9\%),
the change of the order of the polynomial function from two 
to one (0.5\%, 1.2\%),
the tracking efficiency (4\%, 4\%),
{\color{black}{the muon identification efficiency (1.1\%, 1.1\%),
the pion identification efficiency (0.2\%, 0.2\%),}}
{\color{black}{the statistical uncertainty in the efficiency(1.0\%, 0.6\%),}} 
the uncertainty in the $\Upsilon(1S) \rightarrow \mu^+ \mu^-$
decay branching fraction (2.0\%, 2.0\%), and  
the total number of  $\Upsilon(4S)$ events 
(1.3\%, 1.1\%).
The total systematic uncertainty for each data sample
is obtained by adding these contributions in quadrature;
the results are {\color{black}{10.3\%}} and {\color{black}{6.6\%}}
for samples I and II, respectively.
{{\color{black}{The systematic uncertainties from the tracking efficiency, 
$\Upsilon(4S)$ counting, $\mathcal{B}(\Upsilon(1S)\to \mu^+\mu^-)$,
and muon and pion identification efficiencies are treated 
as fully correlated systematic  errors for samples I and II.
   Other uncertainties  are treated as uncorrelated errors.
First the weighted average of the uncorrelated uncertainties 
   $\langle\sigma_{sys}^{uncor}\rangle$ is evaluated.
The total systematic uncertainty is obtained by adding  
$\langle\sigma_{sys}^{uncor}\rangle$  and remaining correlated uncertainties
 in quadrature.}}} 

The measured weighted product branching fraction is \\

$\mathcal{B}(\Upsilon(4S)\to \Upsilon(1S)\pi^+\pi^-)\times
\mathcal{B}(\Upsilon(1S)\to \mu^+\mu^-)$ \\
\centerline{$={\color{black}{(2.11 \pm 0.30(\mathrm{stat.}) 
\pm 0.14(\mathrm{sys.}))}} \times 10^{-6}.$} \\
The branching fraction is \\

$\mathcal{B}(\Upsilon(4S)\to \Upsilon(1S)\pi^+\pi^-)$ \\
\centerline{$=({\color{black}{0.85}} \pm 0.12(\mathrm{stat.}) 
\pm${\color{black}{0.06}}(sys.))
$\times 10^{-4}.$} \\

\noindent We also extract the partial decay width for the
$\Upsilon(4S)\to \Upsilon(1S)\pi^+\pi^-$ transition
using the world-average value of the total
width~\cite{PDG}, and obtain \\

$\Gamma(\Upsilon(4S)\to \Upsilon(1S)\pi^+\pi^-)$ \\
\centerline{ $= {\color{black}{(1.75 \pm 0.25(\mathrm{stat.})
\pm 0.24(\mathrm{sys.}))}}\: \mathrm{keV}.$} \\

\noindent The measured values of
$\mathcal{B}(\Upsilon(4S)\to \Upsilon(1S)\pi^+\pi^-)$ and
$\Gamma(\Upsilon(4S)\to \Upsilon(1S)\pi^+\pi^-)$
supersede our previous results~\cite{4S}  with improved accuracy.
The new Belle results are compatible with those of {\it BABAR}~\cite{etab}.

To summarize, a study of transitions between $\Upsilon$ states
with the emission of charged pions has been performed at Belle.
The peak at $\Delta M \,= \, (1118.7\,\pm\, 1.2)\,\mathrm{MeV}/c^2$ is
interpreted as a signal for the decay
$\Upsilon(4S)\to \Upsilon(1S)\pi^+\pi^-$ 
with a subsequent $ \Upsilon(1S) \rightarrow \mu^+ \mu^-$ transition.
The branching fraction
$\mathcal{B}(\Upsilon(4S)\to \Upsilon(1S)\pi^+\pi^-)$
and the partial decay width
$\Gamma(\Upsilon(4S) \rightarrow \Upsilon(1S) \pi^{+} \pi^{-})$
are measured.
{\color{black}{We have not studied the $\Upsilon(4S)\to \Upsilon(2S)\pi^+\pi^-$
decay because criteria applied to 
the raw experimental data make our sensitivity to this decay limited.}}

We thank the KEKB group for the excellent operation of the
accelerator, the KEK cryogenics group for the efficient
operation of the solenoid, and the KEK computer group and
the National Institute of Informatics for valuable computing
and SINET3 network support.  We acknowledge support from
the Ministry of Education, Culture, Sports, Science, and
Technology (MEXT) of Japan, the Japan Society for the 
Promotion of Science (JSPS), and the Tau-Lepton Physics 
Research Center of Nagoya University; 
the Australian Research Council and the Australian 
Department of Industry, Innovation, Science and Research;
the National Natural Science Foundation of China under
contract No.~10575109, 10775142, 10875115, and 10825524; 
the Department of Science and Technology of India; 
the BK21 program of the Ministry of Education of Korea, 
the CHEP SRC program and Basic Research program (grant 
No. R01-2008-000-10477-0) of the 
Korea Science and Engineering Foundation;
the Polish Ministry of Science and Higher Education;
the Ministry of Education and Science of the Russian
Federation and the Russian Federal Agency for Atomic Energy;
the Slovenian Research Agency;  the Swiss
National Science Foundation; the National Science Council
and the Ministry of Education of Taiwan; and the U.S.\
Department of Energy.
This work is supported by a Grant-in-Aid from MEXT for 
Science Research in a Priority Area (``New Development of 
Flavor Physics''), and from JSPS for Creative Scientific 
Research (``Evolution of Tau-lepton Physics'').


\end{document}